\documentclass[epj]{svjour}
\usepackage{graphics}
\usepackage{psfig}
\newcommand{\bsim}{\mbox{\raisebox{-0.1cm}{$\;
\stackrel{\textstyle>}{\sim}\;$}}}
\newcommand{\lsim}{\mbox{\raisebox{-0.1cm}{$\;
\stackrel{\textstyle<}{\sim}\;$}}}

\begin{document}
\title{Superconductivity of Rb$_3$C$_{60}$:
breakdown of the Migdal-Eliashberg theory} 
\subtitle{}
\author{E.  Cappelluti\inst{1}
\and C. Grimaldi\inst{2} \and L. Pietronero\inst{1}
\and S. Str\"assler\inst{2} \and G.A. Ummarino}
\institute{Dipartimento di Fisica, Universit\'{a} ``La Sapienza", 
P.le A.  Moro 2, 00185 Roma, and INFM Roma1, Italy \\
\and \'Ecole Polytechnique F\'ed\'erale de Lausanne,
D\'epartement de Microtechnique IPM, CH-1015 Lausanne, Switzerland
\and INFM - Dipartimento di Fisica, Politecnico di Torino,
c.so Duca degli Abruzzi 24, 10129 Torino, Italy
}
\date{\today/ \mbox{}}
\abstract{
In this paper, through an exhaustive analysis within the 
Migdal-Eliashberg theory,
we show the incompatibility of experimental data
of Rb$_3$C$_{60}$ with the basic assumptions of the standard
theory of superconductivity.
For different models of the
electron-phonon spectral function $\alpha^2\!F(\Omega)$
we solve numerically the Eliashberg
equations to find which values of the electron-phonon coupling $\lambda$,
of the logarithmic phonon frequency $\Omega_{\rm ln}$ and of the Coulomb
pseudopotential $\mu^*$ reproduce the experimental data of Rb$_3$C$_{60}$.
We find that the solutions are
essentially independent of the particular shape of
$\alpha^2\!F(\Omega)$ and that, to explain the experimental data
of Rb$_3$C$_{60}$, 
one has to resort to extremely large couplings:
$\lambda=3.0\pm 0.8$.
This results differs from the usual partial analyses reported up to now and
we claim that this value exceeds the maximum allowed $\lambda$
compatible with the crystal lattice stability. Moreover, we show
quantitatively that the obtained values of $\lambda$ and $\Omega_{\rm ln}$
strongly violate Migdal's theorem and consequently are incompatible
with the Migdal-Eliashberg theory. One has therefore to consider
the generalization of the theory of superconductivity
in the nonadiabatic regime to account for the experimental properties
of fullerides.
}
\maketitle

\section{Introduction}

In contrast to the cuprates, the other family of high-$T_c$
superconductors, the fullerene compounds, 
shows a quite more conventional phenomenology:
their normal sta\-te \, prop\-er\-ties are Fermi liquid-like,
no stripe formation or signal of pseudogap appear above $T_c$,
they are $s$-wave superconductors with a sizeable carbon 
isotope effect \cite{gunnarsson}. 
It is certainly due to their apparently ordinary 
phenomenology that superconductivity in C$_{60}$ materials 
is now often assumed to be correctly described by the conventional 
Migdal-Eliashberg (ME) theory 
of the electron-phonon driven superconductivity \cite{migdal,elia}.
The relatively high critical temperatures of the 
$A_3$C$_{60}$ fullerene compounds
(up to $T^{\rm max}_c = 40$ K for Cs$_3$C$_{60}$
under pressure \cite{palstra})
are therefore generally thought to be due to an optimized
electron-phonon interaction achieved, according to the different
proposed explanations, by a large electron coupling to
the alkali phonons \cite{zhang}, to the C$_{60}$ rotational modes \cite{mazin1},
or by the so-called
curvature argument for the intramolecular couplings \cite{varma,schluter1}.
According to this latter theory, the electron-phonon coupling 
steadily increases with the curvature of the fulleride molecule so that
compounds with smaller molecular radius (C$_{36}$, C$_{28}$)
are expected to have critical temperatures even higher than those of the
C$_{60}$-based materials \cite{cote}.

The above theories disregard however several aspects of the
phenomenology of the fullerene compounds which do not
fit into the standard ME scenario.
In fact, like the high-$T_c$ copper-oxides, the fullerene compounds
have extremely
low charge carrier densities \cite{uemura}, have significant 
electron correlation \cite{gunnarsson,chakra}, and
are close to a metal-insulator 
transition\cite{iwasa,gkm} showing a strong dependence of $T_c$ upon 
doping \cite{yildirim} and disorder \cite{watson}. From a ME point of
view, these features tend to degrade the superconducting state
so that it appears difficult to understand why the fullerene 
compounds should represent the best optimized ME materials.

In this situation, it should be therefore of primary importance
to assess to which extend the experimental data can be explained 
by the ME theory. This issue has been even more highlighted
by the recent discovery of superconductivity with $T_c =52$ K
in FET hole doped C$_{60}$ \cite{batlogg}.
Until recently, however,
the spread of experimental results has prevented a detailed and definitive
analysis. Nevertheless, in the last years, the experimental uncertainty 
for the alkali fullerene compound Rb$_3$C$_{60}$, which shows the highest 
critical temperature at room pressure
$T_c =30$ K within the A$_3$C$_{60}$ family, has been considerably 
narrowed. 

In fact, resistive measurements on $99$ \% enriched $^{13}$C
single-crystals have recently established the carbon isotope
coefficient $\alpha_{\rm C}$ with the best up-to-date
accuracy, $\alpha_{\rm C}=0.21\pm 0.012$ \cite{fuhrer},
resolving therefore a long standing uncertainty on the value
of $\alpha_{\rm C}$ in fullerides mainly due to
partial isotope substitution and/or magnetic measurements on
powder samples with broad transitions often wider than the isotope
shift itself \cite{gunnarsson,fuhrer}.
In addition, crossed tunneling and optical transmission studies 
on the {\em same} sample have provided
accurate measurements of the zero temperature energy gap $2\Delta$
leading to $2\Delta/T_c=4.2\pm 0.2$ \cite{koller}.
This result is more accurate than previous spectroscopic studies and agrees
with NMR and photoemission measurements \cite{tycko,chun},
moreover it has been obtained by both bulk- (optical transmission)
and surface- (tunneling) sensitive measurements which have
given often contrasting results.

Although we shall briefly consider also
other estimations existing in literature,
in this paper we mainly focus on the above reported experimental data.
The main point in fact is that these data are supplemented by quite 
small error bars, which permit a more rigorous analysis,
We shall show however that different determinations of the data,
in particular of the superconducting gap, would even more substain
the results presented in this paper.

The experimental values above discussed
seem, at first sight, perfectly compatible with the ME theory.
For example, $T_c=30$ K is not far from
$T_c=23.2$ K of Nb$_3$Ge and $2\Delta/T_c=4.2\pm 0.2$ is very close
to $4.18$ and $4.25$ of V$_3$Ga and Nb$_3$Al(2), respectively \cite{carbotte}.
Finally, the carbon isotope coefficient $\alpha_{\rm C}=0.21$ is
similar to the isotope effects of the elemental ME superconductors
Os ($\alpha=0.2$) and Mo ($\alpha=0.33$) \cite{parks}.
The apparently ordinary values of $T_c$, $\alpha_{\rm C}$, and 
$2\Delta/T_c$, independently considered, could therefore
be used as arguments in favour of the validity of the ME theory 
for Rb$_3$C$_{60}$. 
However, each of this values has little meaning if taken individually.
For example, in Ref. \cite{koller} the experimental data
$T_c=30$ K and $2\Delta/T_c=4.2$, but not $\alpha_{\rm C}=0.21$, have
been fitted by setting $\lambda=1.16$, $\mu^*=0.1$, and $\Omega_{\rm ln}=302$ K,
while in Ref. \cite{fuhrer} $T_c=30$ K and $\alpha_{\rm C}=0.21$,
but not $2\Delta/T_c=4.2$, have been fitted by $\lambda=0.9$, $\mu^*=0.22$,
and $\Omega_{\rm ln}=1360$ K. Although the discrepancies in the values
of $\lambda$ and $\mu^*$ are somehow acceptable, the two values of 
$\Omega_{\rm ln}$ differ by a factor of five.

A previous partial analysis, based only on the values of
the critical temperature $T_c=30$ K
and of the isotope coefficient $\alpha_{\rm C}=0.21$, has pointed
out an intrisic inconsistency of the ME framework with respect
of the adiabatic assumption \cite{cgps}.
In this paper we extend this study by taking into account
also the estimation of the superconducting gap $2\Delta/T_c=4.2$
and by considering realistic electron-phonon interactions related
both to inter- and intra- molecular modes. We show that the
experimental data of Rb$_3$C$_{60}$, when
analyzed in the context of the fullerene compounds, $T_c=30$ K, 
$\alpha_{\rm C}=0.21$ and $2\Delta/T_c=4.2$ are actually
incompatible with the standard ME framework. We accomplish this
task in Sec. \ref{numerical} by considering different models of the electron-phonon 
interaction and by numerically solving the ME equation in order 
to reproduce the experimental data. We find  in Sec. \ref{discussion} 
that the resulting values of the electron-phonon 
interaction are always far too large to avoid lattice instabilities
and to neglect the electron-phonon vertex corrections beyond Migdal's
limit, as required by the ME formulation \cite{migdal,elia}.
Finally, in Sec. \ref{concl} we propose that the main origin of
the failure of the ME framework lies in the breakdown of the 
adiabatic hypothesis (Migdal's theorem) which, in doped fullerenes, 
is naturally driven by the small value of the Fermi energy.

\section{The Migdal-Eliashberg equations}
\label{MEequations}

As pointed out in the introduction, the superconducting state of
the fullerene compounds is quite often regarded in terms of the
ME theory. Since the alkali doped fullerenes are three dimensional
$s$-wave superconductors, the superconducting properties are therefore 
thought to be correctly described by the standard ME 
equations \cite{elia,carbotte}:
\begin{eqnarray}
Z(i\omega_n)&=& 1+\frac{\pi T}{\omega_n}
\sum_m \int_0^{\infty} d\Omega 
\frac{\alpha^2\!F(\Omega)\: 2\Omega}{\Omega^2+(\omega_n-\omega_m)^2}
\nonumber\\
&&\hspace{3cm}\times
\frac{\omega_m}{\sqrt{\omega_m^2+\Delta(i\omega_m)}},
\label{ee1}
\end{eqnarray}
\begin{eqnarray}
Z(i\omega_n)\Delta(i\omega_n)&=&\pi T
\sum_m\Bigg[
\int_0^{\infty} d\Omega 
\frac{\alpha^2\!F(\Omega)\: 2\Omega}{\Omega^2+(\omega_n-\omega_m)^2}
\nonumber\\
&&
-\mu\: \theta(\omega_c-|\omega_m|)\Bigg]
\frac{\Delta(i\omega_m)}{\sqrt{\omega_m^2+\Delta(i\omega_m)}}.
\label{ee2}
\end{eqnarray}
In the above equations $Z(i\omega_n)=1-\Sigma(i\omega_n)/i\omega_n$,
where $\Sigma(i\omega_n)$ is the normal state electronic self-energy,
$\Delta(i\omega)$ is the Matsubara gap-function and $\omega_n$ and
$\omega_m$ are Matsubara fermionic frequencies. 

In the above equations, $\alpha^2F\!(\Omega)$ is the electron-phonon
spectral function (also known as the Eliashberg function)
which defines the electron-phonon coupling constant $\lambda$ and
the logarithmic frequency $\Omega_{\rm ln}$ through the following 
relations \cite{carbotte}:
\begin{equation}
\label{lambda}
\lambda=2\int\frac{d\Omega}{\Omega} \alpha^2\!F(\Omega),
\end{equation}
\begin{equation}
\label{omegaln}
\ln\Omega_{{\rm ln}}=\frac{2}{\lambda} 
\int \frac{d\Omega}{\Omega}\ln(\Omega) \alpha^2\!F(\Omega). 
\end{equation}
The dimensionless parameter $\mu$ represents the effective
Coulomb repulsion probed by the Cooper pair
at the  energy scale $\omega_c$ which is
much larger than the phonon energy scale. 
It is clear that the value
of the Coulomb parameter $\mu$ depends on the specific value of
$\omega_c$ which is somewhat arbitrary \cite{combescot}.
In fact a more sound quantity is the pseudopotential $\mu^*$ defined as:
\begin{equation}
\mu^*=\frac{\mu}{1+\mu\ln\left(\omega_c/\Omega_{\rm max}\right)},
\label{mu-mu*}
\end{equation}
where $\Omega_{\rm max}$ is the maximum phonon frequency.
The physical properties of the superconducting state depend on
$\mu^*$ rather than $\mu$ or $\omega_c$ \cite{combescot}. 
In the numerical solution of Eqs. (\ref{ee1})-(\ref{ee2}) we
have set $\omega_c=5\Omega_{\rm max}$. Larger values of $\omega_c$ do
not modify the results and we have checked that different choises of
$\omega_c$, although providing different values of $\mu$ as expected,
lead to the same $\mu^*$ via Eq. (\ref{mu-mu*}).

The quantities we want to extract from the ME equations
(\ref{ee1}) and (\ref{ee2}) are the critical temperature $T_c$, the
zero temperature superconducting gap $\Delta$ and the carbon
isotope coefficient $\alpha_{\rm C}$. To achieve this, we must first choose
the input quantities $\alpha^2\!F(\Omega)$ and $\mu$ (or $\mu^*$).
Of course, the electron-phonon coupling $\lambda$ and the 
logarithmic phonon frequency $\Omega_{\rm ln}$ follows directly from
a given Eliashberg function $\alpha^2\!F(\Omega)$ 
via Eqs. (\ref{lambda})-(\ref{omegaln}). Our aim is to find for which values
of $\lambda$, $\Omega_{\rm ln}$ and $\mu^*$ the ME equations have
$T_c=30$ K, $\alpha_{\rm C}=0.21$, and $2\Delta/T_c=4.2\pm 0.2$ as solutions. 
We solve numerically Eqs. (\ref{ee1})-(\ref{ee2}) to obtain
the critical temperature $T_c$ and the zero-temperature
``Matsubara'' gap 
$\Delta_0=\lim_{T\rightarrow 0}\Delta(i\omega_{n=0})$.

\begin{figure}
\centerline{\psfig{figure=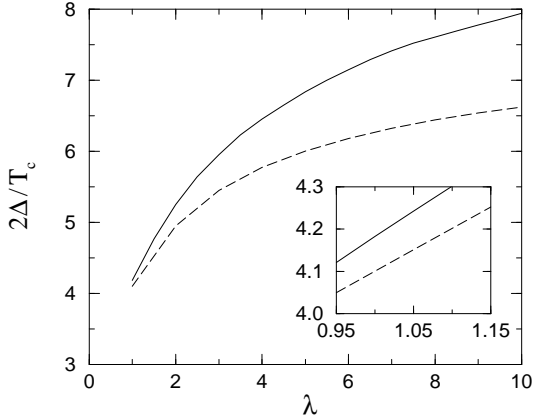,width=7cm}}
\caption{Plot of $2\Delta/T_c$ (solid line) and $2\Delta_0/T_c$
(dashed line) as function of $\lambda$ for an Einstein phonon spectrum
and $\mu=0$. In inset, a zoom of the region around $2\Delta/T_c=4.2$.}
\label{figrealax}
\end{figure}

The physical gap $\Delta(T)$
can be obtained from $\Delta(i\omega_n)$ via the analytical continuation 
on the real-axis\cite{marsiglio} and through the relation
\begin{equation}
\Delta(T)=\mbox{Re}\left[\Delta(\omega=\Delta(T),T)\right].
\label{gapph}
\end{equation}
In order to quantify the discrepancy
between $\Delta$ and $\Delta_0$, in Fig. \ref{figrealax}
the physical and Matsubara gaps are plotted as function of $\lambda$
for a Einstein phonon model with $\mu=0$.
The enhancement of $\Delta$ with respect to $\Delta_0$
is essentially driven by the size of the gap itself.
An enlargement of the region relevant for Rb$_3$C$_{60}$
($2\Delta/T_c \simeq 4.2$)
is shown in inset of Fig. \ref{figrealax}.
The discrepancy is
of order of $2\%$, from $2\Delta_0/T_c \simeq 4.10$ to 
$2\Delta/T_c \simeq 4.18$. The total effect is in any case less than the
experimental error of $\Delta$ in Rb$_3$C$_{60}$ \cite{koller}.
Numerical solutions of Eliashberg
equations in imaginary-axis therefore provide a quite good
determination even of the physical gap $\Delta$ within
the experimental accuracy available for Rb$_3$C$_{60}$.
Note that the above discussion holds true even in the presence
of Coulomb repulsion and for generic $\alpha^2\!F(\Omega)$.
As we have pointed out, discrepancies between $\Delta_0$ and $\Delta$
are essentially related to the size of $\Delta$, and therefore
to $2\Delta/T_c$. The physical gap can be quite small, leading to
weak-intermediate coupling phenomenology, even for large values
of $\lambda$ when the electron-phonon coupling is balanced by
a strong Coulomb repulsion, as is the case for fullerene 
compounds \cite{gunnarsson}.

Finally, the carbon isotope coefficient $\alpha_{\rm C}$, defined by
\begin{equation}
\alpha_{\rm C}=-\frac{M_{\rm C}}{\Delta M_{\rm C}}\frac{\Delta T_c}{T_c},
\label{carbon}
\end{equation}
is numerically evaluated by solving the
ME equations (\ref{ee1}) and (\ref{ee2})
for two values of the
carbon mass: $M_{\rm C}$ and $M_{\rm C}+\Delta M_{\rm C}$.
For a single component material, 
it can be shown that isotope
substitution enters the ME equations only through
a scaling factor in the frequency dependence of the Eliashberg
function, namely \cite{carbotte,rainer}:
\begin{equation}
\alpha^2\!F(\Omega) \equiv {\cal F}\left(\sqrt{M} \Omega\right),
\label{isot1}
\end{equation}
where $M$ is the element mass. In a two-component system, 
like Rb$_3$C$_{60}$, the electron-phonon spectral function
$\alpha^2\!F(\Omega)$ contains in principle mixed phonon modes
involving C$-$C, Rb$-$Rb and C$-$Rb displacements.
The corresponding contributions to $\alpha^2\!F(\Omega)$ scale
therefore in different ways with isotope substitution, and one should
deal with partial isotope effects \cite{carbotte,rainer}.
However,  inter-mol\-ec\-u\-lar modes involving the alkali ions appear to couple
negligibly to electrons in A$_3$C$_{60}$ systems.
Evidence for this conclusion comes from the negligible
effect on $T_c$ upon isotope substitution of the alkali ions \cite{ebbesen},
implying that the electron-phonon spectral
function $\alpha^2\!F(\Omega)$ does not contain Rb-phonon modes.
In addition, the dependence of $T_c$ on the lattice parameter
is identical both by applying pressure and by chemical substitution
of the alkali atoms \cite{fleming}. This result points out the marginal role
played by alkali atoms on superconductivity: they mainly
tune the lattice constant by interspacing the buckyballs molecules
and provide charge carriers in the conduction band. But they
do not effectively couple to electrons.
Based on this experimental evidence, we assume the Eliashberg
function $\alpha^2\!F(\Omega)$ is determined only by carbon modes.
In this case, using of Eq. (\ref{isot1}) is perfectly justified.

\section{Numerical analysis}
\label{numerical}

We are now in the position to analyze the experimental situation
of Rb$_3$C$_{60}$.
We solve numerically the ME equations
for different shapes of $\alpha^2\!F(\Omega)$ 
with the constraints given by the measured data of Rb$_3$C$_{60}$.
For the Eliashberg function $\alpha^2\!F(\Omega)$ we use
i) a single $\delta$-peak (Einstein spectrum, model I),
ii) a broad spectrum defined by a rectangular function (model II), and
iii) a broad (rectangular) spectrum with an additional
low frequency contribution describing the coupling to the
inter-molecular modes (model III).

\subsection{Model I: Einstein phonon spectrum}
\label{modelI}

\begin{figure}
\centerline{\psfig{figure=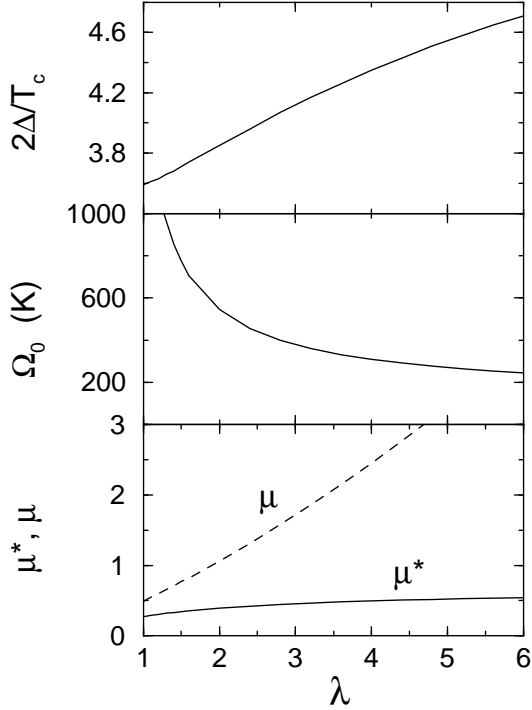,width=7cm}}
\caption{Plot of $2\Delta/T_c$, $\Omega_0$, $\mu$ and $\mu^*$ as
functions of $\lambda$ as obtained by
the numerical solutions of Eqs. (\ref{ee1})-(\ref{ee2}) with an Einstein
phonon spectrum and with the
conditions $T_c=30$ K and $\alpha_{\rm C}=0.21$.}
\label{Einstein}
\end{figure}

As a particularly simple but representative case
we first consider a $\delta$-peaked spectrum 
describing a single Einstein phonon with frequency $\Omega_0$
and electron-phonon coupling constant $\lambda$:
\begin{equation}
\label{delta-peak}
\alpha^2\!F(\Omega)=\frac{\lambda\Omega_0}{2}\delta(\Omega-\Omega_0).
\end{equation}
The use of a single $\delta$-function $\alpha^2\!F(\Omega)$ makes
the ME equations to be characterized just by the three microscopic
parameters $\lambda$, $\Omega_0$ and $\mu$. 
Once the critical temperature $T_c=30$ K and the isotope coefficient
$\alpha_{\rm C}=0.21$ are fixed, this permits to obtain a one-to-one correspondence
between the value of the ratio $2\Delta/T_c$ and a set of $\lambda$, 
$\Omega_0$, $\mu$. The numerical results are shown in Fig. \ref{Einstein}
where we plot also the pseudopotential $\mu^*$ obtained from the 
calculated $\mu$ via Eq. (\ref{mu-mu*}). 

The experimental ratio $2\Delta/T_c = 4.2\pm 0.2$, together with $T_c=30$ K
and $\alpha_{\rm C}=0.21$, is obtained by an electron-phonon coupling
$\lambda=3.33^{+0.90}_{-0.79}$, an Einstein phonon 
$\Omega_0=\Omega_{\rm ln}=350^{+84}_{-52}$ K,
bare and screened Coulomb repulsions respectively $\mu=1.95^{+0.68}_{-0.54}$
and $\mu^*=0.47^{+0.03}_{-0.04}$,
where the error bars result from the experimental uncertainty 
of $2\Delta/T_c$ \cite{noteerr}.
We note that the obtained value of $\Omega_{\rm ln}$ is compatible with
the low-frequency intra-molecular modes\cite{hebard} and that 
$\mu^*\sim 0.4-0.5$
is close to the most accurated theoretical estimations \cite{zwick,koch}. 
However, $\lambda=3.33^{+0.90}_{-0.79}$ largely exceeds $\lambda\simeq 1$
which is the estimatation most commonly found in literature \cite{gunnarsson}.
More importantly, as we show later, the value we have found is too large 
to prevent lattice instabilities and to ensure the validity of Migdal's theorem.
This result does not rely on the specific shape
of $\alpha^2\!F(\Omega)$.
As we show below, different shapes of $\alpha^2\!F(\Omega)$
lead to similar results in terms of $\lambda$, $\Omega_{\rm ln}$
and $\mu^*$.

\subsection{Model II: rectangular phonon spectrum}
\label{modelII}

Let us consider now the case of a spectral function with a finite width.
To this end,
we schematize $\alpha^2\!F(\Omega)$ as
a rectangle centered at $\Omega_0$ and having width $\Delta\Omega_0$.
By using Eq. (\ref{lambda}), $\alpha^2\!F(\Omega)$ can therefore be
written as:
\begin{equation}
\label{alfamodelII}
\alpha^2\!F(\Omega)= 
\frac{\lambda\theta(\Omega-\Omega_0+\Delta\Omega_0/2)
\theta(\Omega_0+\Delta\Omega_0/2-\Omega)}
{2\ln|(\Omega_0+\Delta\Omega_0/2)/(\Omega_0-\Delta\Omega_0/2)|},
\end{equation}
where $\theta$ is the Heaviside step function.
The quantity $\Delta\Omega_0/\Omega_0$ parametrizes 
the finite width: for $\Delta\Omega_0/\Omega_0 \rightarrow 0$
the single Einstein phonon case (model I) is recovered, while the limit
$\Delta\Omega_0/\Omega_0 \rightarrow 2$ corresponds to a constant
$\alpha^2\!F(\Omega)$ ranging from 0 to $2\Omega_0$.
This latter case should be considered merely as a mathematical limit with no
physical relevance since, as $\Omega\rightarrow 0$, $\alpha^2\!F(\Omega)$
should always vanish.
The characteristic quantities $\lambda$ and $\Omega_{\rm ln}$
are determined as usual from Eqs. (\ref{lambda}) and (\ref{omegaln}).
Note that $\Omega_{\rm ln}$ does not coincide with $\Omega_0$ as in 
the Einstein model, but it is actually given by:
\begin{equation}
\label{omegalnI}
\Omega_{\rm ln}=\left[\Omega_0^2-
\left(\frac{\Delta\Omega_0}{2}\right)^2\right]^{1/2}.
\end{equation}
Hence, $\Omega_{\rm ln}$ is always smaller than $\Omega_0$
and the equality 
$\Omega_{\rm ln}=\Omega_0$ holds true only in the limit
$\Delta\Omega_0/\Omega_0 \rightarrow 0$.

\begin{figure}
\centerline{\psfig{figure=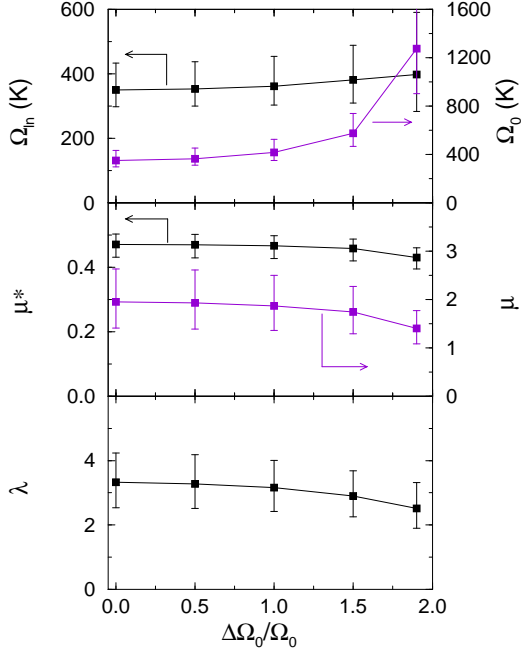,width=7cm}}
\caption{Plot of $\lambda$, $\mu^*$, $\Omega_{\rm ln}$ (left side axes)
and $\Omega_0$ and $\mu$ (right side axes) as function of the broadening
of the phonon spectrum $\Delta\Omega_0/\Omega_0$. 
$T_c$, $\alpha$ and $2\Delta/T_c$ are fixed to their experimental values.}
\label{figbroad}
\end{figure}

The effects of the finite width of the electron-phonon spectrum
are shown in Fig. \ref{figbroad}, where the quantities
$\lambda$, $\mu^*$, $\Omega_{\rm ln}$, as well as
$\Omega_0$ and $\mu$, obtained by requiring
$T_c=30$ K, $\alpha_{\rm C}=0.21$, $2\Delta/T_c=4.2\pm 0.2$,
are plotted as function of the parameter $\Delta\Omega_0/\Omega_0$.
The error bars correspond to the experimental uncertainty on
$2\Delta/T_c$.
From the analysis of Fig. \ref{figbroad} some important remarks
can be stated. First of all, we see that 
$\lambda$, $\mu^*$ and $\Omega_{\rm ln}$ have only
a weak dependence on the phonon spectrum width: this sustains
the idea that $T_c$, $\alpha_{\rm C}$, and $2\Delta/T_c$
depend essentially only on the ``McMillan'' parameters
$\lambda$, $\mu^*$ and $\Omega_{\rm ln}$ regardless of the particular
shape of $\alpha^2\!F(\Omega)$. 
Moreover, it shows also a lower bound for $\lambda$ ($\lambda > 2$), which,
although lower than the Einstein phonon case, is still outside
the range of validity of Migdal-Eliashberg theory.
Note that values of $\Delta\Omega_0/\Omega_0 > 1.9$ represent solutions
with phonon spectra exceeding $\sim 2300$ K, which is
the largest intramolecular phonon frequency \cite{gunnarsson,hebard},
so that they are solutions not compatible with the real materials.

Further results can be deduced from Fig. \ref{figbroad}. In fact,
intramolecular phonons in Rb$_3$C$_{60}$ extends in a range of energies 
$393$ K$ < \Omega < 2266$ K \cite{gunnarsson,hebard}.
If we assume that all these phonons couple
with equal strength to the electrons, then this situation would correspond
to $\Omega_0 \simeq 1330$ K and $\Delta\Omega_0/\Omega_0 \simeq 1.4$. 
However, from Fig. \ref{figbroad}, $\Delta\Omega_0/\Omega_0 = 1.4$ corresponds
to a phonon spectrum centered at $\Omega_0 \simeq 550$ K, which is much
lower than $\Omega_0 \simeq 1330$ K. Furthermore, if we keep only
$T_c=30$ K and $\alpha_{\rm C}=0.21$ fixed, then for $\Omega_0 = 1330$ K 
and $\Delta\Omega_0/\Omega_0 = 1.4$ we find $2\Delta/T_c \simeq 3.7$.
Again, this limiting
situation is incompatible with the experimental data of Rb$_3$C$_{60}$.

\subsection{Model III: rectangular spectrum
plus inter-molecular modes}
\label{modelIII}

In the past, the issue of determining if intermolecular or
intramolecular phonon modes are more responsible for the
superconductivity phenomenon in fullerides has been widely debated.
Although an active role of alkali-C$_{60}$ phon\-ons has been ruled
out by the zero alkali isotope effect \cite{ebbesen},
intermolecular buckyball modes can in principle play a not negligible role
in the electron-phonon coupling.
Indeed, numerical calculations indicate that the intramolecular phonons
couple relatively strongly to the conduction 
electrons \cite{varma,schluter1,schluter2,faulhaber,antropov,breda}, but
a contribution from very low-frequency intermolecular modes (librations)
has been claimed to provide a better fit to some experimental 
da\-ta \cite{mazin1}. In literature, there are examples of
numerical calculations which favour\cite{chen} or 
disfavour\cite{zwick,picket,antropov} an important
contribution of the intermolecular modes to the total $\lambda$.

In this section we consider the effects of an eventual intermolecular
contribution to the total electron-phonon coupling. We schematize
this contribution by adding a low-frequency part to the rectangular
model of $\alpha^2\!F(\Omega)$ studied above. We do not distinguish
between libration (of typical energy $50$ K \cite{mazin1}),
and acoustic C$_{60}$-C$_{60}$ modes 
(of frequency up to $\sim 90-100$ K \cite{wang}),
and treat all the intermolecular couplings as a continuum ranging
from $\Omega=0$ to $\Omega=\Omega_{\rm inter}$, where we set
$\Omega_{\rm inter}= 100$ K as the maximum intermolecular frequency.
The overall shape of this low-frequency part is of secondary importance,
the only condition is that it should vanish as $\Omega\rightarrow 0$
[this ensures that the intermolecular contribution of $\lambda$ is
finite, see Eq. (\ref{lambda})]. We have therefore chosen the following
expression for the total $\alpha^2\!F(\Omega)$ : 
\begin{equation}
\label{alfamodelIII}
\alpha^2\!F(\Omega)= \left\{
\begin{array}{lrcl}
A_{\rm inter}\, \Omega &  0 & \leq \Omega \leq &\Omega_{\rm inter}\\
A_{\rm intra}          & 
\Omega_0-\frac{\displaystyle \Delta\Omega_0}{\displaystyle 2}& 
\leq \Omega \leq & 
\Omega_0+\frac{\displaystyle \Delta\Omega_0}{\displaystyle 2}
\end{array}
\right.,
\end{equation}
where, from Eq. (\ref{lambda}), $A_{\rm inter}$ and $A_{\rm intra}$ 
are related respectively 
to the intermolecular and intramolecular electron-phonon coupling constants,
$\lambda_{\rm inter}$ and $\lambda_{\rm intra}$, as:
\begin{equation}
\label{ainter}
A_{\rm inter}=\frac{\lambda_{\rm inter}}{2\,\Omega_{\rm inter}},
\end{equation}
and
\begin{equation}
\label{aintra}
A_{\rm intra}=\frac{\lambda_{\rm intra}}
{2\ln|(\Omega_0+\Delta\Omega_0/2)/(\Omega_0-\Delta\Omega_0/2)|}.
\end{equation}
Moreover, form Eq. (\ref{omegaln}), the logarithmic frequency is:
\begin{equation}
\label{OlnmodelIII}
\Omega_{\rm ln}=\left(\frac{\Omega_{\rm inter}}
{\rm e}\right)^{\frac{\displaystyle \lambda_{\rm inter}}{\displaystyle \lambda}}
\left[\Omega_0^2-\left(\frac{\Delta\Omega_0}{2}\right)^2\right]^
{\frac{\displaystyle \lambda_{\rm intra}}{\displaystyle 2\lambda}},
\end{equation}
where $\lambda=\lambda_{\rm inter}+\lambda_{\rm intra}$ is the
total electron-phonon coupling. 
In this model, the coupling to the intermolecular phonons is
modulated just by $\lambda_{\rm inter}$, while $\Omega_{\rm inter}$
is kept fixed at $100$ K. Of course,
for $\lambda_{\rm inter}=0$ the present model reduces to the previous 
model II were only intramolecular modes are taken into account.

\begin{figure}[t]
\centerline{\psfig{figure=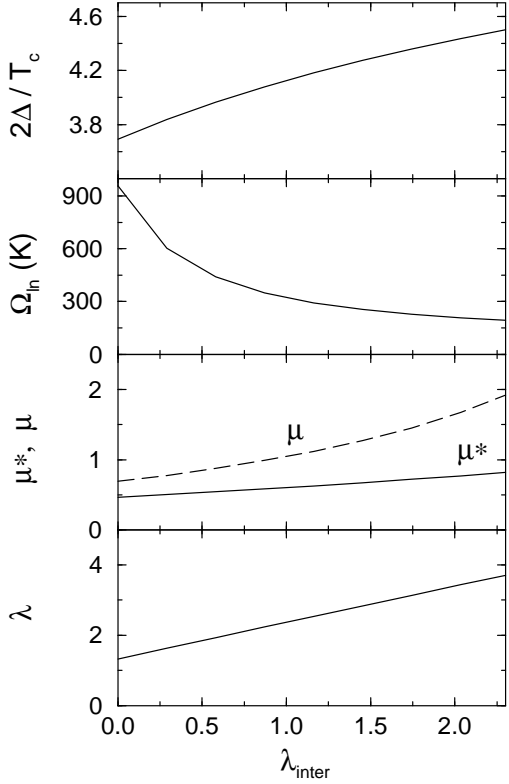,width=7cm}}
\caption{Dependence of $\lambda$, $\mu$, $\mu^*$, 
$\Omega_{\rm ln}$ and $2\Delta/T_c$ on the intermolecular electron-phonon
coupling $\lambda_{\rm inter}$ keeping fixed
$T_c=30$ K and $\alpha=0.21$.}
\label{figtwofixed}
\end{figure}

Let us start by considering the case in which all the intramolecular
phonons ($393$ K$<\Omega < 2266$ K, which corresponds to
$\Omega_0=1330$ K and $\Delta\Omega_0/\Omega_0=1.4$)
couple with the same weight
to the electrons. We have seen before that for this limiting case
there is no solution compatible with the experimental data of
Rb$_3$C$_{60}$.
By switching on the intermolecular electron-phonon coupling, however,
it is now possible to find a solution which reproduces all the experimental
values of $T_c$, $\alpha_c$ and $2\Delta/T_c$.
The behaviour of $2\Delta/T_c$, together with $\Omega_{\rm ln}$, $\mu$, $\mu^*$,
and the total electron-phonon coupling $\lambda$,
for fixed  $T_c=30$ K and $\alpha_{\rm C}=0.21$, as function of 
$\lambda_{\rm inter}$ is plotted in Fig. \ref{figtwofixed}. 
For $\lambda_{\rm inter}=0$ we again find $2\Delta/T_c\simeq 3.7$, while
$2\Delta/T_c=4.2\pm 0.2$ is obtained for 
$\lambda_{\rm inter}=1.47^{+0.80}_{-0.63}$ which corresponds to
$\lambda = 2.85^{+0.83}_{-0.65}$, $\mu=0.90^{+0.14}_{-0.10}$,
$\mu^* =0.37^{+0.02}_{-0.02}$ and $\Omega_{\rm ln} = 178^{+99}_{-50}$ K.
Although now a solution exists for $\lambda_{\rm inter}\neq 0$,
the corresponding values of $\lambda$, $\Omega_{\rm ln}$, 
and $\mu^*$ are of the same order of those extracted by the previous model I
and model II.

\begin{figure}[t]
\centerline{\psfig{figure=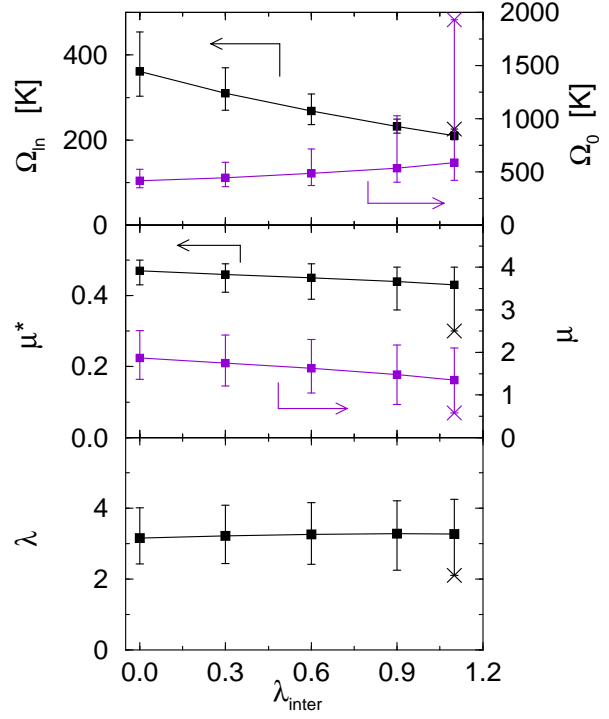,width=8cm}}
\caption{Dependence of $\lambda$, $\mu$, $\mu^*$, 
$\Omega_{\rm ln}$ and $\Omega_0$ on the intermolecular electron-phonon
coupling $\lambda_{\rm inter}$. All the data have been obtained by
requiring $T_c=30$ K, $\alpha_{\rm C}=0.21$, and $2\Delta/T_c=4.2\pm 0.2$
with the exception of the results for $\lambda_{\rm inter}=1.1$ where
the lowest value of $2\Delta/T_c$ has been set equal 
to $4.04$ (marked by a cross)}
\label{figtwomove}
\end{figure}

Given $\lambda_{\rm inter}\neq 0$, we have also studied the effect of the
broadening of the intramolecular modes by using different values of
$\Delta\Omega_0$. For a given $\Delta\Omega_0/\Omega_0$,
the frequency $\Omega_0$ is adjusted as function of $\lambda_{\rm inter}$
to reproduce $T_c=30$ K, $\alpha_{\rm C}=0.21$, and $2\Delta/T_c=4.2 \pm 0.2$. 
The results
for $\Delta\Omega_0/\Omega_0=1$ are shown in Fig. \ref{figtwomove}. 
The smaller range of intramolecular phonons leads to solutions for all
intermolecular couplings up to $\lambda_{\rm inter}=0.9$. 
For $\lambda_{\rm inter}=1.1$ the error bars of $\Omega_0$ becomes
so large that the maximum phonon frequency $\Omega_{\rm max}=
\Omega_0+\Delta\Omega_0/2$ exceeds the highest possible intramolecular
phonon energy ($2266$ K). In particular, for $2\Delta/T_c=4.04$ we have obtained
$\Omega_0=1930$ K (marked by a cross in Fig. \ref{figtwomove}) which 
corresponds to $\Omega_{\rm max}=2895$ K. The value of $\Omega_0$
for $2\Delta/T_c=4.0$ would be even higher but too computing demanding
to evaluate exactly.

For $\Delta\Omega_0/\Omega_0=0$ (Einstein phonon plus
intermolecular contribution) there are solutions also at higher
values of $\lambda_{\rm inter}$, but for $\lambda_{\rm inter}< 0.9$ the
resulting $\lambda$, $\mu^*$, and $\Omega_{\rm ln}$ nearly
exactly overlap with those of Fig. \ref{figtwomove}. Hence,
compared to the case in which all the intramolecular phonons partecipate
to the coupling ($\Delta\Omega_0/\Omega_0=1.4$ and $\Omega_0=1330$ K,
for which we have found $\lambda=2.85^{+0.83}_{-0.65}$), lower values
of $\Delta\Omega_0/\Omega_0$ tend to give higher values of $\lambda$.

\section{Breakdown of the adiabatic hypothesis}
\label{discussion}

\begin{table}[b]
\caption{Summary of the numerical solutions of the ME equations
for $T_c=30$ K, $\alpha_{\rm C}=0.21$ and $2\Delta/T_c=4.2\pm 0.2$.
For Model II and Model III we report only the set of values which
defines lower limits of $\lambda$.}
\begin{tabular}{l|c|c|c}
&&& \\
 & $\Omega_{\rm ln} [K]$ & $\mu^*$ & $\lambda$ \\
&&& \\
\hline \hline
&&&\\
Model I & $350^{+84}_{-52}$ & $0.47^{+0.03}_{-0.04}$ & $3.33^{+0.90}_{-0.79}$ \\
&&&\\
\hline
&&&\\
Model II & $<398^{+193}_{-115}$ & $>0.43^{+0.03}_{-0.04}$ & $>2.51^{+0.81}_{-0.61}$ \\
&&&\\
\hline
&&&\\
Model III & $>178^{+99}_{-50}$ & $>0.37^{+0.02}_{-0.02}$ & $>2.85^{+0.83}_{-0.65}$ \\
&&&\\
\end{tabular}
\label{tablesumm}
\end{table}

We summarize in Table \ref{tablesumm} the main results obtained by solving the
ME equations (\ref{ee1}) and (\ref{ee2}) under the different models of the
electron-phonon spectral function $\alpha^2\! F(\Omega)$ defined in 
Secs. \ref{modelI}-\ref{modelIII}. 
For model II and model III
we report the results which give only the lower values
of $\lambda$ ($\Delta\Omega_0/\Omega_0=1.9$ for model II and 
$\Delta\Omega_0/\Omega_0=1.4$, $\Omega_0=1330$ K, and 
$\lambda_{\rm inter}=1.47^{+0.8}_{-0.63}$ for model III). 

The main result which can be extracted from Table \ref{tablesumm} is that,
independently of the particular form of $\alpha^2\! F(\Omega)$ considered,
the experimental data of Rb$_3$C$_{60}$ can be solutions of the ME
equations only for very large values of $\lambda$. Considering that
model I and the lower limits of model II and model III reported in Table \ref{tablesumm}
correspond to quite extreme situations, we estimate $\lambda\simeq 3\pm 0.8$
as the most plausible value for more realistic shapes of $\alpha^2\! F(\Omega)$.

Let us examine now the consequences of this result and the consistency
with the whole ME framework. A first crucial point is that such
high values of the electron-phonon coupling exceed the maximum allowed 
$\lambda$ compatible with the lattice stability. Perturbative calculations
of the phonon propagator show that
for $\lambda\rightarrow 1$ the phonon frequencies at small momenta
are renormalized to zero \cite{AGD}, while recent calculations on the Holstein model
predict the breakdown of the ME expansion at $\lambda>1.25$ \cite{pata1}.
For real materials, it is argued that $\lambda\simeq 1.5$ is a good
estimate of the maximum electron-phonon coupling compatible with
lattice stability \cite{anderson}.
Note moreover that, even for $\lambda$ well below the 
criterion of lattice stability, the system could undergo other kinds
of instabilities, like charge-density-waves, which would prevent
superconductivity.

In addition to the problem concerning the lattice stability, the
results obtained in the last section, when analyzed together with
the band structure of Rb$_3$C$_{60}$, lead to a serious inconsistency
with respect to the validity of Migdal's theorem \cite{migdal}.
The use
of Equations (\ref{ee1}) and (\ref{ee2}) implicitly assumes the adiabatic
hypothesis according to which $E_{\rm F}\gg \Omega_{\rm ph}$,
where $\Omega_{\rm ph}$ is the typical phonon frequency and $E_{\rm F}$
is the Fermi energy. In general, the adiabatic hypothesis ensures that the
electron-phonon vertex corrections, absent in the ME equation, can be
neglected \cite{migdal}. In fact, according to Migdal, the order of magnitude of the
vertex corrections is at least
\begin{equation}
\label{P1}
P=2\int\!d\Omega\,\frac{\alpha^2\!F(\Omega)}{E_{\rm F}}=
\lambda\frac{\Omega_{\rm ph}}{E_{\rm F}},
\end{equation}
where  $\Omega_{\rm ph}=
\frac{2}{\lambda}\int\!d\Omega\,\alpha^2\!F(\Omega)$ \cite{noteomega}.
Note however that the adiabatic condition $E_{\rm F}\gg \Omega_{\rm ph}$ 
does not automatically give $P\ll 1$ since $P\sim 1$ could be obtained for
$\lambda\gg 1$. Hence, our results are consistent with the ME theory
if both $\Omega_{\rm ph}/E_{\rm F}$ and $P$ are negligible.
\begin{figure}
\centerline{\psfig{figure=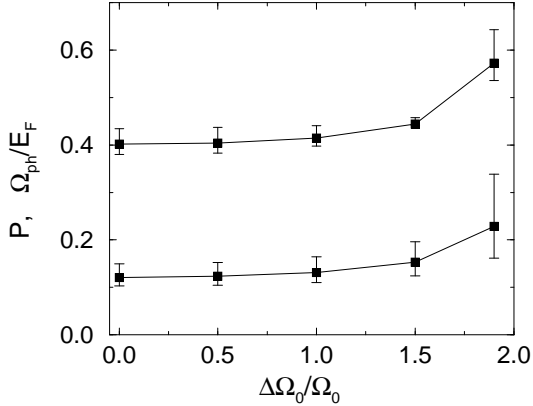,width=7cm}}
\caption{Adiabatic ratio $\Omega_{\rm ph}/E_{\rm F}$ (lower line)
and Migdal's parameter $P=\lambda \Omega_{\rm ph}/E_{\rm F}$
(upper line) as function of the broadening of the phonon spectrum
in model II.}
\label{figmigII}
\end{figure}
\begin{figure}
\centerline{\psfig{figure=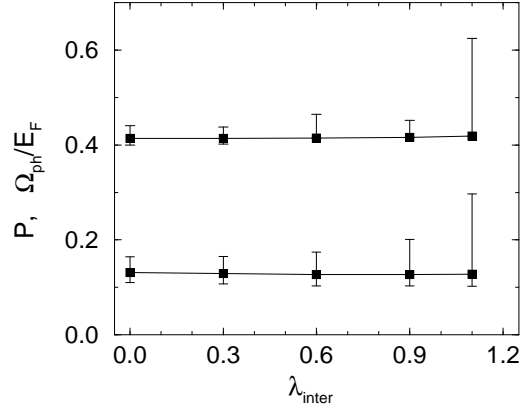,width=7cm}}
\caption{Adiabatic ratio $\Omega_{\rm ph}/E_{\rm F}$ (lower line)
and Migdal's parameter $P=\lambda \Omega_{\rm ph}/E_{\rm F} $
(upper line) as function of the intermolecular electron-phonon coupling
in model III ($\Delta\Omega_0/\Omega_0=1.0$).}
\label{figmigIII}
\end{figure}

The electronic band structure of the fullerides is given by a set of
very narrow subbands of width $W\sim 0.5$ eV 
separated from each other 
by gaps of order $U \sim 0.5$ eV or larger \cite{gunnarsson}.
For the A$_3$C$_{60}$ compounds
the conduction $t_{\rm 1u}$ subband is half-filled by electrons.
Due to coupling to the phonons, the conduction electrons
experience both intra- and inter-band scatterings.
However, the inter-band electron-phonon couplings have only
a negligible effect on superconductivity because they involve
transitions of order $W+U\simeq 1$ eV or larger \cite{varma}.
The relevant bandwidth is thus that of the $t_{\rm 1u}$ conduction
band and the corresponding Fermi energy is
$E_{\rm F} = W/2 \simeq 0.25$  eV $= 2900$ K.
This value should be compared with the characteristic phonon
frequency scale.

According to (\ref{P1}), in the case of an Einstein phonon spectrum (model I)
the average phonon frequency
$\Omega_{\rm ph}$ reduces to $\Omega_0$. From the previous results 
of Sec. \ref{modelI}
we obtain therefore an adiabatic ratio
$\Omega_{\rm ph}/E_{\rm F}=0.12^{+0.03}_{-0.02}$, which is
moderately nonadiabatic.
However, the large val\-ue of $\lambda$ reported in Table \ref{tablesumm}
leads to an important contribution of the vertex corrections,
as estimated by Eq. (\ref{P1}),
with a magnitude $P=0.40^{+0.03}_{-0.02}$ far from to
be negligible.
The effects on $P$ and $\Omega_{\rm ph}/E_{\rm F}$
of the broadening (model II) and of the inclusion of inter-molecular 
modes (model III) in $\alpha^2\!F(\Omega)$ are reported 
in Fig. \ref{figmigII} and \ref{figmigIII},
respectively.
For model II, the explicit expression of $P$ can be 
obtained from Eqs. (\ref{alfamodelII}) and (\ref{P1}) and reduces to:
\begin{equation}
\label{P2}
P=\frac{\lambda}{\ln\left(\frac{\displaystyle \Omega_0+\Delta\Omega_0/2}
{\displaystyle \Omega_0-\Delta\Omega_0/2}\right)}
\frac{\displaystyle \Delta\Omega_0}{\displaystyle E_{\rm F}},
\,\,\, \,\,\,\,\,\,(\mbox{model II})
\end{equation}
while for model III, Eq. (\ref{alfamodelIII}), $P$ becomes:
\begin{equation}
\label{P3}
P=\frac{\lambda_{\rm inter}}{2}\frac{\Omega_{\rm inter}}{E_{\rm F}}+
\frac{\lambda-\lambda_{\rm inter}}{\ln\left(\frac{\displaystyle \Omega_0+\Delta\Omega_0/2}
{\displaystyle \Omega_0-\Delta\Omega_0/2}\right)}\frac{\Delta\Omega_0}{E_{\rm F}}.
\,\,\, \,\,\,(\mbox{model III})
\end{equation}

From Fig. \ref{figmigII} it is apparent that the broadening affects only weakly
the Einsteing phonon results, $\Delta\Omega_0/\Omega_0 =0$,
at least for $\Delta\Omega_0/\Omega_0\leq 1$. For larger values of
$\Delta\Omega_0/\Omega_0$ both $P$ and $\Omega_{\rm ph}/E_{\rm F}$
increase signalling an even stronger violation of the adiabatic hypothesis
and of Migdal's theorem.
Switching on the low frequency inter-molecular couplings, 
Fig. \ref{figmigIII},
does not modify appreciably the results of Fig. \ref{figmigII}.
In fact, the effect of 
non zero values of $\lambda_{\rm inter}$ is to shift $\Omega_0$ to slightly
higher frequencies than for $\lambda_{\rm inter}=0$ 
(see Fig. \ref{figtwomove}), 
leading to an almost $\lambda_{\rm inter}$-independent $\Omega_{\rm ph}$.

Summarizing the results obtained for $E_{\rm F}=0.25$ eV, 
we conclude that $\Omega_{\rm ph}/E_{\rm F}\bsim 0.12$ and $P\bsim 0.4$
independently of the shape of $\alpha^2\!F(\Omega)$. Hence,
the experimental data of Rb$_3$C$_{60}$ are not consistent
with the ME theory since both the adiabatic
hypothesis, $\Omega_{\rm ph}/E_{\rm F}\ll 1$, and Migdal's theorem,
$P\ll 1$, are violated. 

How much this result depends on the precise determination of
the superconducting gap size? All over our analysis we have
consider an experimental uncertainty $2\Delta/T_c=4.2\pm 0.2$ and
we have shown that within this error bars any attempt to describe
the experimental data  breaks down the adiabatic
hypothesis. However, recent measurements by surface-sensitive techniques
seem to point towards a BCS-like value of the reduced gap
$2\Delta/T_c\simeq 3.53$ \cite{hesper}, and one could question the validity
of our conclusions for such a small value of $\Delta$.
However, the violation of ME theory is even stronger in this case.
As shown in Fig. \ref{Einstein} for an Einstein
phonon spectrum (model I), BCS-like values of $\Delta$ need coupling with
extremely high phonon energies in order to
reproduce $T_c=30$ K and $\alpha_{\rm C}=0.21$. In particular,
$2\Delta/T_c = 3.6$ implies $\Omega_0=1350$ K,
corresponding to an adiabatic ratio 
$\Omega_{\rm ph}/E_{\rm F} \simeq 0.46$ and to a Migdal's parameter
$P \simeq 0.49$. 
The failure of ME theory in even more drastic if
broader spectra (models II and III) are considered.
Indeed, as discussed in Sec. \ref{numerical}, in these cases a lowest value 
$2\Delta/T_c \simeq 3.7$ is obtained and a BCS-like
gap $2\Delta/T_c = 3.53$, together with $T_c=30$ K and $\alpha_{\rm C}=0.21$,
is incompatible with the ME framework.

\section{Discussion and conclusions}
\label{concl}

The critical analysis of the experimental data carried on in the previous sections
has pointed out the inconsistency of the standard ME theory for Rb$_3$C$_{60}$.
We address now the origin of such inconsistency in the hypothesis that
phonons are still the mediators of superconductivity in Rb$_3$C$_{60}$.
To this end, let us consider some existing theoretical calculations of the
coupling of the $t_{\rm 1u}$
electrons to the $H_{\rm g}$ intra-molecular phonon modes. Each phonon
mode has frequency $\Omega_i$, $i=1,\ldots,8$, and couples to 
the $t_{\rm 1u}$ electrons via $V_i=\lambda_i/N_0$,
$N_0$ being the electron density of states per spin at the Fermi level.
\begin{figure}
\protect
\centerline{\psfig{figure=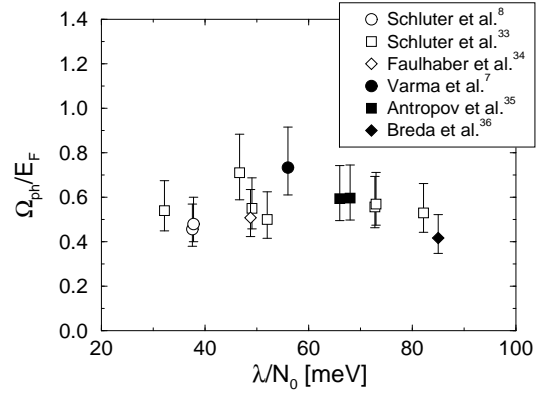,width=7cm}}
\caption{Adiabatic parameter $\Omega_{\rm ph}/E_{\rm F}$ 
extracted from var\-i\-ous calculations of the intramolecular
electron-phonon interaction in fullerenes.}
\label{figLDA}
\end{figure}
In Fig. \ref{figLDA} we show a collection of data taken from various 
calculations \cite{varma,schluter1,schluter2,faulhaber,antropov,breda},
and for each set of data
we have estimated the corresponding adiabatic parameter 
$\Omega_{\rm ph}/E_{\rm F}$ where, from Eq. (\ref{P1}), 
\begin{equation}
\label{omega_teo}
\Omega_{\rm ph}=\frac{2}{\lambda}\int \! d\Omega\,\alpha^2\!F(\Omega)=
\frac{1}{V}\sum_{i=1}^{8} \Omega_i V_i,
\end{equation}
and $V=\sum_iV_i$.
The data refer to different
schemes including tight-binding, LDA,  ab-initio etc., 
and the error bars stem from the uncertainty in the calculated $t_{\rm 1u}$
bandwidth (we have assumed $E_{\rm F}=0.25\pm 0.05$ eV).

The important point of Fig. \ref{figLDA} is that, despite of the large spread 
of the values of $\lambda/N_0$,
all these calculations agree in estimating the adiabatic parameter 
$\Omega_{\rm ph}/E_{\rm F}$ to be larger than about $0.4$, so that
Migdal's theorem breaks down and the whole ME framework is 
invalidated. Of course, the reason for such high values of $\Omega_{\rm ph}/E_{\rm F}$
stems from the fact that, independently of details, the $H_{\rm g}$ phon\-ons 
have energies ranging from $30$ meV to $200$ meV while the conduction 
$t_{\rm 1u}$ electron band has a width of only $W=0.4-0.6$ eV.

With the exception of few cases (see for example
Ref. \cite{antropov}) the problem
concerning the non-validity of the adiabatic hypothesis is ignored or,
at best, underestimated and the numerical results are often
claimed to provide a strong evidence that, after all, the fullerene
compounds are standard ME superconductors \cite{notefuhrer}.
Our opinion is instead
that such numerical calculations show quite clearly an evident nonadiabaticity
of the electron-phonon interaction for which the ME framework is
completely inadeguate. Hence, a correct description of the superconducting
state in fullerides should be formulated in such a way
that $\Omega_{\rm ph}/E_{\rm F}$ can assume values larger than zero
for which the electron-phonon vertex corrections, as well as
finite-band effects, must be included from the start.
As by-products of the very small value of $E_{\rm F}$, also non-constant
density of state effects and strong electron correlations may play a role in the
superconducting properties of fullerides. However, these effects alone
can hardly explain the high values of $T_c$ while, under favourable
circumstances, the electron-phonon vertex corrections can 
effectively enhance the pairing interaction \cite{psg}, leading
to an increase of $T_c$ for values of $\lambda$ lower than those 
needed in the ME theory \cite{pata2}.

It would be therefore interesting to test whether and for which parameter
values a theory generalized beyond Migdal's theorem can describe
the experimental data of Rb$_3$C$_{60}$. The accomplishement of this task
is outside the scope of this work. However
as shown in Ref. \cite{cgps}, preliminary results suggest a
positive role of the nonadiabatic electron-phonon effects.
For example, while in the adiabatic ME equations 
a $\lambda=1-4$  is needed in order to
reproduce $T_c=30$ K and 
$\alpha_{\rm C}=0.21$ for realistic values of phonon frequencies
$300$ $K$ $ \lsim \Omega_{\rm ph} \lsim 1500-1800$ K, a much more
reasonable $\lambda <1$ is needed in the nonadiabatis theory
of superconductivity where first nonadiabatic corrections
are taken into account \cite{cgps}.

These results are quite interesting since they suggest
that the experimental data of Rb$_3$C$_{60}$ could be explained by
values of $\lambda$ much closer to those of the intercalated compounds
of graphite ($\lambda\sim 0.3$). This is an important remark because the intramolecular
phonons in C$_{60}$ stem from the carbon-carbon bonds just as the highest
phonon modes of graphite. In this perspective, the large discrepancy
between the superconducting critical temperature
in graphite intercalated compounds
($T_c \sim 0.2$ K) and in fullerides 
would come from the difference of the electronic structures. 
Graphite compounds have large Fermi energies locating these materials
in the adiabatic regime where ME equations actually hold, while the
small values of $E_{\rm F}$ of the fullerides give rise to the opening
of nonadiabatic channels enhancing the pairing.


\begin{thebibliography}{99}
\bibitem{noteerr}
Here and in the following, we neglect the error
bars in $\alpha_{\rm C}$ since they add only a small fraction
of the uncertainty of $2\Delta/T_c$ to the values of
$\lambda$, $\Omega_{\rm ln}$, and $\mu^*$.
\bibitem{noteomega}
Note that in Eq. (\ref{P1}) enters the averaged phonon frequency
$\Omega_{\rm ph}$ rather than the logarithmic frequency
$\Omega_{\rm ln}$. They coincide in case of a single 
$\delta$-peak spectrum, 
while in general $\Omega_{\rm ln}\leq\Omega_{\rm ph}$.
\end{thebibliography}
\end{document}